\documentclass[journal]{IEEEtran}

\usepackage{myStyleIEEE}
\usepackage{nomencl}
\IEEEoverridecommandlockouts
\newtheorem*{remark}{Remark}

\usepackage{etoolbox}
\renewcommand\nomgroup[1]{%
  \item[\bfseries
  \ifstrequal{#1}{P}{A. Parameters}{%
  \ifstrequal{#1}{V}{C. Variables}{%
  \ifstrequal{#1}{S}{B. Sets and Indices}{}}}%
]}

\begin{document}
\bstctlcite{IEEEexample:BSTcontrol}

\title{Short Circuit Current Constrained UC in High IBG-Penetrated Power Systems}

\newtheorem{proposition}{Proposition}
\renewcommand{\theenumi}{\alph{enumi}}

\newcommand{\uros}[1]{\textcolor{magenta}{$\xrightarrow[]{\text{Uros}}$ #1}}

\author{Zhongda~Chu,~\IEEEmembership{Student~Member,~IEEE,} 
        Fei~Teng,~\IEEEmembership{Member,~IEEE} 
        
\vspace{-0.5cm}}
\maketitle
\IEEEpeerreviewmaketitle

\begin{abstract}
Inverter Based Generators (IBGs) have been increasing significantly in power systems. Due to the demanding thermal rating of Power Electronics (PE), their contribution to the system Short Circuit Current (SCC) is much less than that from the conventional Synchronous Generators (SGs) thus reducing the system strength and posing challenges to system protection and stability. This paper proposes a Unit Commitment (UC) model with SCC constraint in high IBG-penetrated systems to ensure minimum operation cost while maintaining the SCC level at each bus in the system. The SCC from synchronous generators as well as the IBGs are explicitly modeled in the formulation leading to an SCC constraint involving decision-dependent matrix inverse. This highly nonlinear constraint is further reformulated into linear form conservatively. The influence of the SCC constraint on the system operation and its interaction with the frequency regulation are demonstrated through simulations on IEEE 30- and 118-bus systems. 
\end{abstract}


\makenomenclature
\mbox{}
\nomenclature[P]{$J$}{Lumped inertia of WT driven systems$\,[\mathrm{Mkgm}^2]$}

\section{Introduction} \label{sec:1}
Concerns on power system operation and stability have been raised with the increasing penetration of Renewable Energy Sources (RESs), which are connected to the grid through power electronic inverters. One of the most significant impact of these IGBs is the decrease of system inertia, which would lead to larger frequency RoCoF and deviation during system disturbances, such as the 9 Aug 2019 London blackout \cite{9Aug}. Extensive research has been carried out to investigate the negative impact on system inertia reduction with various solutions being proposed. 

However, fewer attention has been paid on another influence associated with the high IBG penetration in the system, i.e., shortage of Short Circuit Current (SCC). Unlike the conventional synchronous generators which are able to provide a short circuit current of 5-10 p.u.\cite{TLEIS2019597}, the IBGs only supply fault current of approximate 1-2 p.u. \cite{7878663} due to the limited overloading capability of the semiconductors. Therefore, as the IBGs in the system continue to increase, the SCC is about to reduce. 
It has been predicted by the National Grid that a decline of 15\% on the average GB short circuit level will be seen by the year 2025 with the largest declined area being identified as North-East and Midlands regions \cite{NGESO_1}. Similarly, the SCC in Germany power system is expected to decrease by up to 50\% at certain buses \cite{Kubis}.

The insufficient SCC in the system would have negative impact on various aspects. Protection devices may take longer or even be not able to detect faults in the system, thus posing risks to system operation and stability \cite{AUEMO}. In addition, the inadequacy in SCC may lead to unsettled voltage oscillations in the network and would further cause lose of synchronization of the Phase-Locked Loop (PLL) \cite{NGESO_2}.

Attempts in various respects have been proposed to enhance the SCC and address the issues it brings in a high IBG-penetrated system. A method to improve the semiconductor's transient thermal capacity and over-current capacity is discussed in \cite{9025185} and the experiments show that an SCC of 3 p.u. for 3 s or 1.5 p.u. for 30 s is feasible. \textcolor{black}{This is achieved by inserting phase-change material right under the chips and setting the phase-change temperature slightly below the maximum permissible chip temperature. However, additional manufacture cost is unavoidable and the effectiveness may be limited without wide applications.} On the other hand, \textcolor{black}{it is demonstrated in \cite{Dimitrios20} that distance protection will be compromised in terms of successful detection and response time as the IBGs' penetration approaches 100\%.} It is also shown that the deployment and control of synchronous condensers can help to mitigate the decline of the short circuit level and its impact on the protection system. An optimal synchronous condenser allocation approach is presented in \cite{Jia18} to maintain the system minimum SCC. \textcolor{black}{However, the mixed-integer nonlinear based problem formulation and} the considerable investment cost (up to $1 \mathrm{M\$}$ fixed cost plus $3 \mathrm{M\$}$ per $100\mathrm{MVar}$) may limit its large scale deployment. In the operational timescale, \cite{Gu19} proposes a system strength and inertia constrained optimal generator dispatch model, where explicit constraints are developed to achieve the optimal system scheduling while satisfying the SCC requirement. \textcolor{black}{However, since the nonlinear SCC constraint is linearized by neglecting the interaction between different generators, an iterative process is necessary to determine the optimal scale factor for SCC calibration} and the contribution from IBGs is not considered.

In fact, due to the significantly distinguished fault characteristic from the conventional synchronous generators, the quantification of the fault current from the IBGs is complicated. \textcolor{black}{A method considering the impact of reinforcement on voltage dip required by the grid code is proposed in \cite{6304664} where the IBGs are modeled as voltage sources behind an impedance.}This equivalent impedance is iteratively updated to match the terminal voltage. \textcolor{black}{The IEC 60909-0-2016 standard \cite{IEC} models the IBGs as constant current source and  approximates the SCC by adding the contribution from IBGs to the classic SCC expression yet neglecting the pre-fault operating condition, which may lead to overestimation of the SCC in the system. \cite{Gu19} proposes a fault current iterative solver, which properly models the current limitation and voltage control logic of IBGs.} However, all these methods either require iterations or are too complex to obtain an analytical expression, thus unable to be directly incorporated into a UC problem.

In order to alleviate the influence of the SCC reduction in a high IBG-penetrated system, an SCC constrained UC model is proposed in this paper. The total number of online generators and their locations are optimized such that the system operation cost is minimized while maintaining the SCC at each bus within a predefined range specified by system operators. The contribution of this paper is identified as:
\begin{enumerate}
    \item A novel SCC constraint is proposed by applying the superposition principle to calculate the SCC in a high IBG-penetrated system, which is then effectively linearized through data-driven method while maintaining both conservativeness and accuracy.
    
    \item A Stochastic UC (SUC) model with SCC and frequency constraints is proposed and formulated as Mixed-Integer Linear Programming (MILP). The SUC model, for the first time, allows to optimize the system operation with simultaneous consideration of variability, uncertainty, low inertia capability and low SCC contribution from IBGs.

    \item The effectiveness and the scalability of the model are assessed through IEEE 30- and 118-bus systems. The results illustrate the interactions between SCC and frequency constraints and highlight the necessity to consider SCC limit in the system scheduling once frequency support from IGBs becomes available. In addition, the case study demonstrate the potential benefit of over-sizing or expanding short-term PE overloading capability of IBGs.
\end{enumerate}

\textcolor{black}{The rest of the paper is organized as follows. Section II introduces the classic SCC calculation method from SGs first based on which a modified approach is proposed to include the contribution from IBGs. A SCC constrained SUC model is formulated in Section III with the highly nonlinear SCC constraint being effectively linearized. Simulation results are provided in Section IV demonstrating the effectiveness of the proposed method. Finally, Section V concludes the paper.}

\section{Short Circuit Analysis} \label{sec:2}
This section first revises the conventional SCC calculation method where the synchronous generators are considered as the only sources of the SCC in the system. An approach incorporating the SCC contribution from the IBGs is then proposed based upon this. Note that only the three-phase nodal short circuit faults are considered in this paper.

\subsection{Classic Short Circuit Current Calculation} \label{sec:2.1}
\begin{figure}[!b]
  \centering
    \begin{minipage}{0.5\textwidth}
        \centering
        \scalebox{0.68}{\includegraphics[trim={1.3cm 0cm 1.0cm .0cm}]{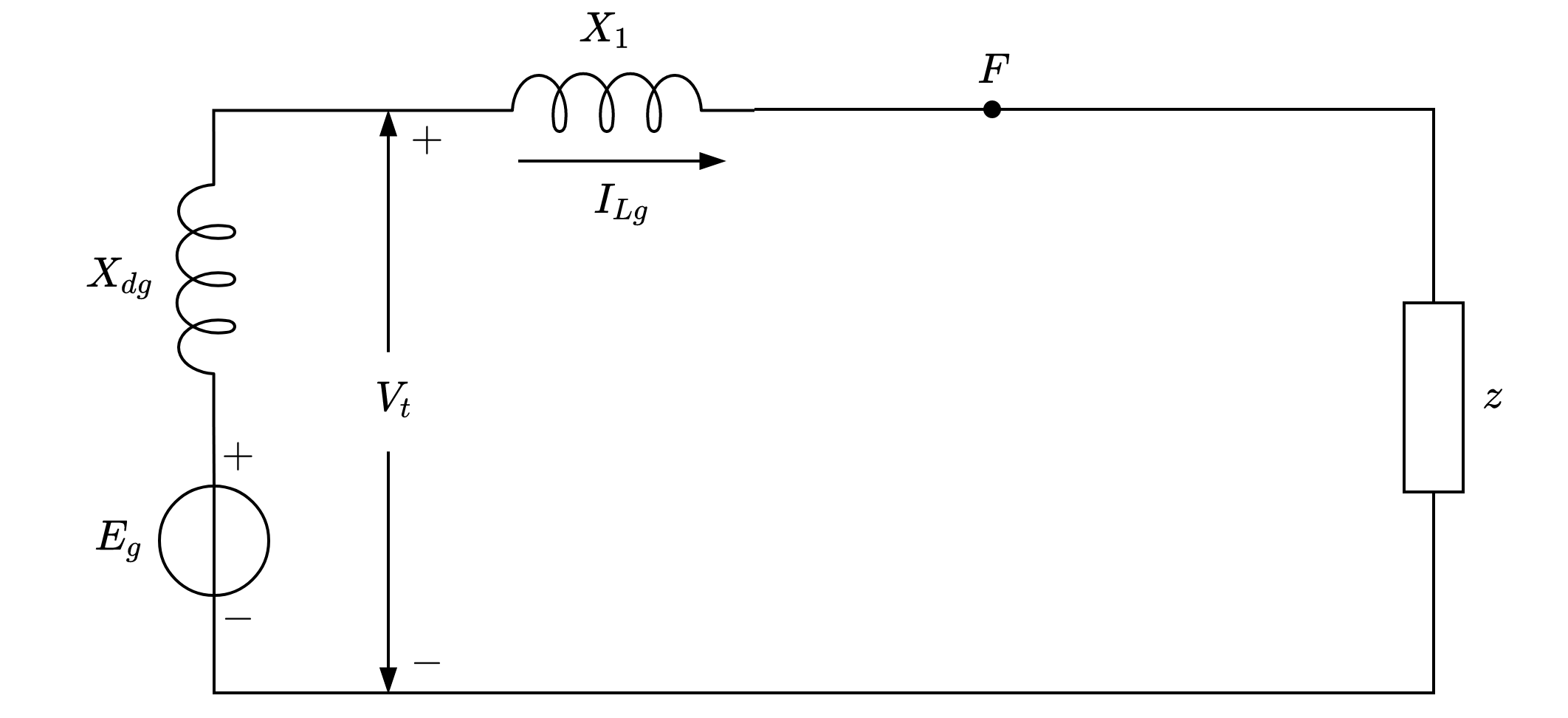}} 
        \vspace{0.15cm}
    \end{minipage} 
    \begin{minipage}{0.5\textwidth}
        \centering
        \scalebox{0.68}{\includegraphics[trim={1.3cm 0cm 1.0cm .0cm}]{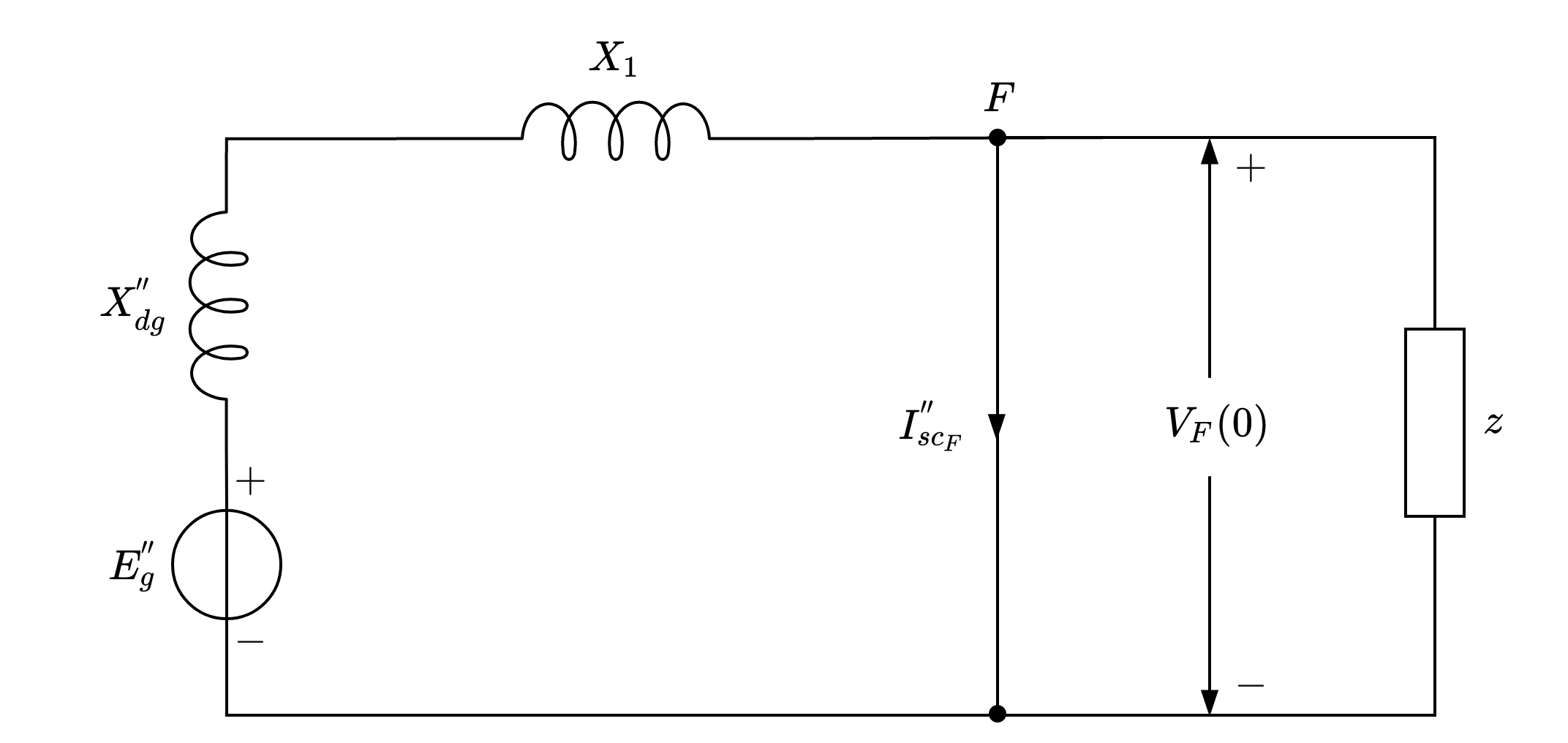}}
    \end{minipage} 
  \caption{\label{fig:SG}Synchronous generator model for SCC computation: (i) pre-fault equivalent circuit; (ii) circuit for $I_{sc_F}^{''}$ calculation.}
\end{figure}
In a conventional power system with low IBG penetration, the SCC at the fault bus mostly comes from the synchronous machines, which can be modeled as an internal voltage in series with a reactance \cite{saadat2010power}. Fig.~\ref{fig:SG} shows the equivalent circuit of a SG, $g\in \mathcal{G}$ with a balanced three-phase load $z$. $X_1$ is the external reactance between the SG and the fault location F. During normal operation as in Fig.~\ref{fig:SG}-(i), the load current $I_{Lg}$ is supplied by the SG internal voltage $E_g$ in series with synchronous reactance $X_{dg}$ leading to a terminal voltage $V_t$. If a three-phase fault occurs at point $F$, Fig.~\ref{fig:SG}-(ii) is used to calculate the subtransient SCC $I_{sc_F}^{''}$, which is supplied by the subtransient internal voltage $E_g^{''}$ in series with subtransient reactance $X_{dg}^{''}$. $E_g^{''}$ is determined by the pre-fault terminal voltage and the load current as follows \cite{grainger1994power}:
\begin{equation}
\label{SG1}
    E_g^{''} = V_t + \mathrm{j} X_{dg}^{''} I_{Lg}.
\end{equation}
The voltage source $E_g^{''}$ in series with a reactance $X_{dg}^{''}$ is further converted to a current source $I_g$ in parallel with a susceptance $Y_g$ according to the Norton's theorem:
\begin{align}
    I_g & = \frac{E_g^{''}}{X_{dg}^{''}} \label{Ig} \\
    Y_g & = \frac{1}{\mathrm{j} X_{dg}^{''}}.
\end{align}
The short circuit current at different buses in the system can be then computed based on the classic superposition method \cite{grainger1994power}. The pre-fault operating conditions are calculated first through power flow analysis to determine the pre-fault voltage $V_{F}(0)$ at the fault bus $F$. The sources $I_g$ of SGs are the only current injections in the system whereas the load currents are ignored at this stage. The voltage $V_{F}(0)$ is then applied in an inverse direction at the short circuit bus while all the other sources in the network are disregarded (open circuit for the current sources) to represent the pure-fault situation. Due to the linear modeling of system elements, the post-fault system can be represented by a linear combination of the pre- and pure-fault conditions. As a result, the initial subtransient short circuit current of the F-th bus $I_{SC_F}^{''}$ can be determined by:
\begin{equation}
    I_{SC_F}^{''} = \frac{-V_{F}(0)}{Z_{FF}}
\end{equation}
where $Z_{FF}$ is the F-th element of the main diagonal in the system impedance matrix $Z$. It should be noted that the subtransient reactance of the SGs are also included in the $Z$ matrix formulation.

\subsection{Modeling of SCC with IBGs} \label{sec:2.2}
With the increasing penetration of IBGs, most of the grid codes require the provision of reactive current from those units to the grid during fault conditions to support the system. \textcolor{black}{However, due to the limited SCC capacity of Voltage Source Converters (VSCs) and the manufacturer-specific control strategies, it is difficult to formulate an accurate method to quantify the SCC contribution from the IBGs on a system level especially with a high IBG penetration.}
\begin{figure}[!t]
  \centering
    \begin{minipage}{0.5\textwidth}
        \centering
        \scalebox{0.68}{\includegraphics[trim={0.3cm 0cm 1.0cm .0cm}]{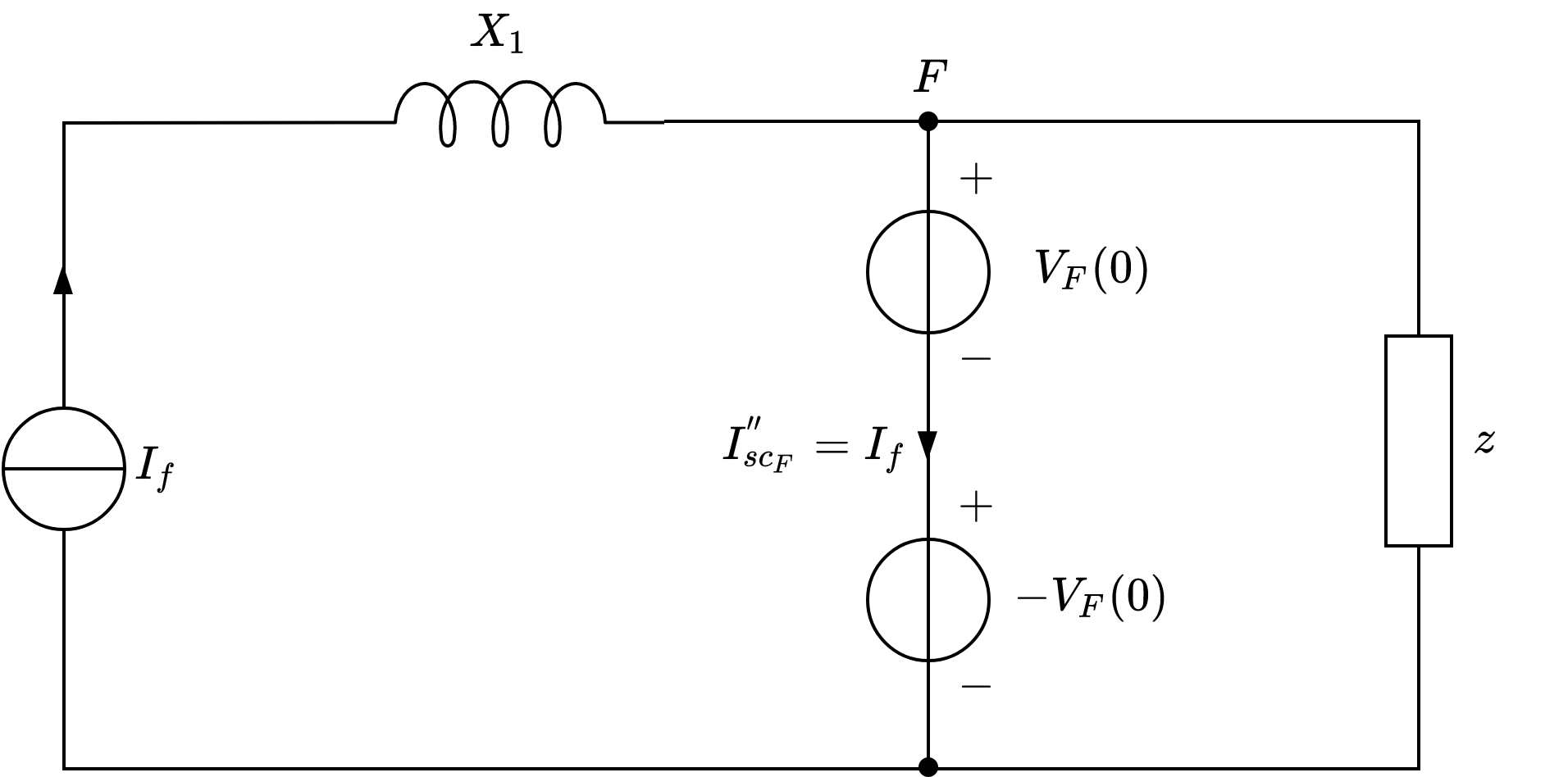}} 
        \vspace{0.15cm}
    \end{minipage} 
    \begin{minipage}{0.5\textwidth}
        \centering
        \scalebox{0.68}{\includegraphics[trim={0.3cm 0cm 0.3cm .0cm}]{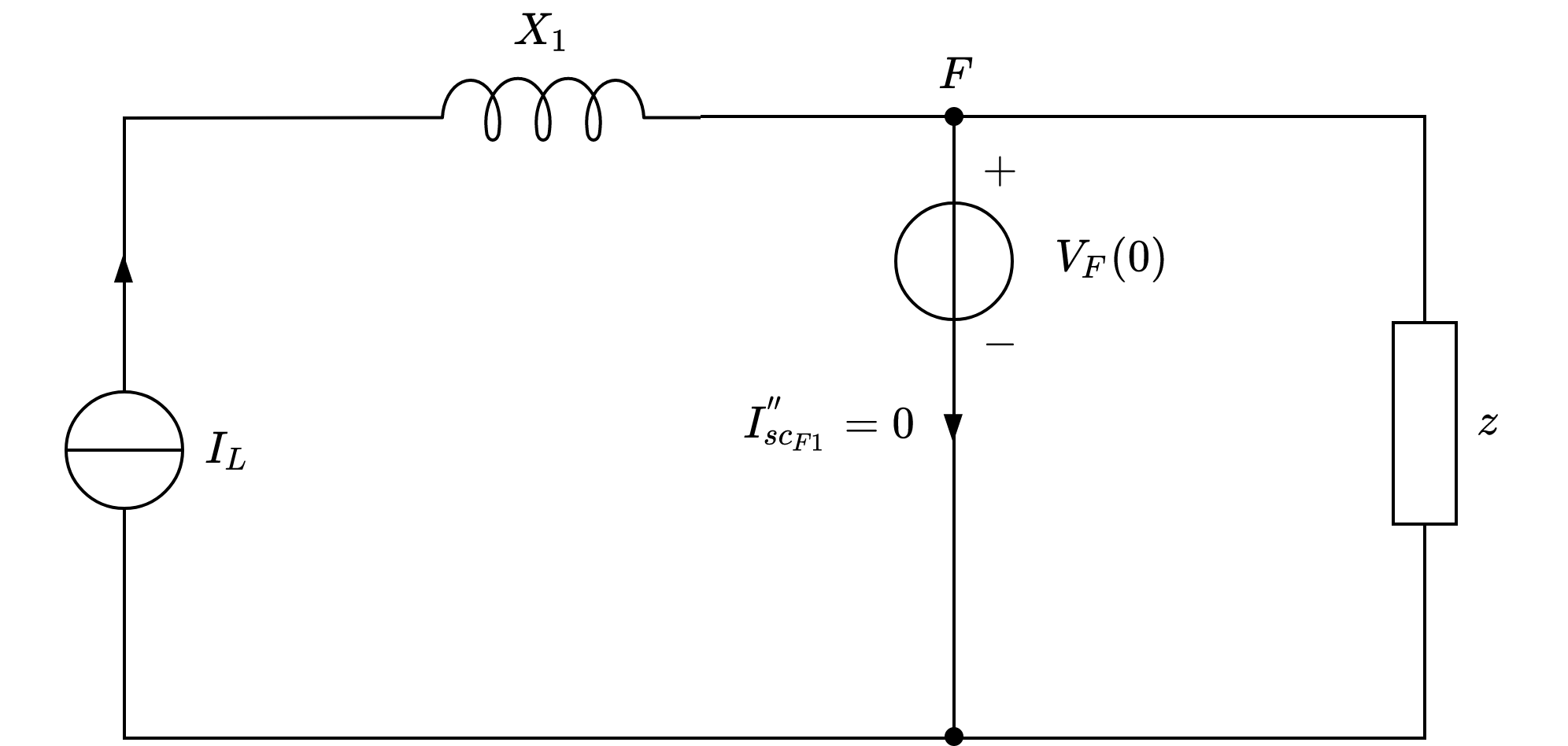}}
    \end{minipage} 
    \begin{minipage}{0.5\textwidth}
        \centering
        \scalebox{0.68}{\includegraphics[trim={0.3cm 0cm 0.8cm .0cm}]{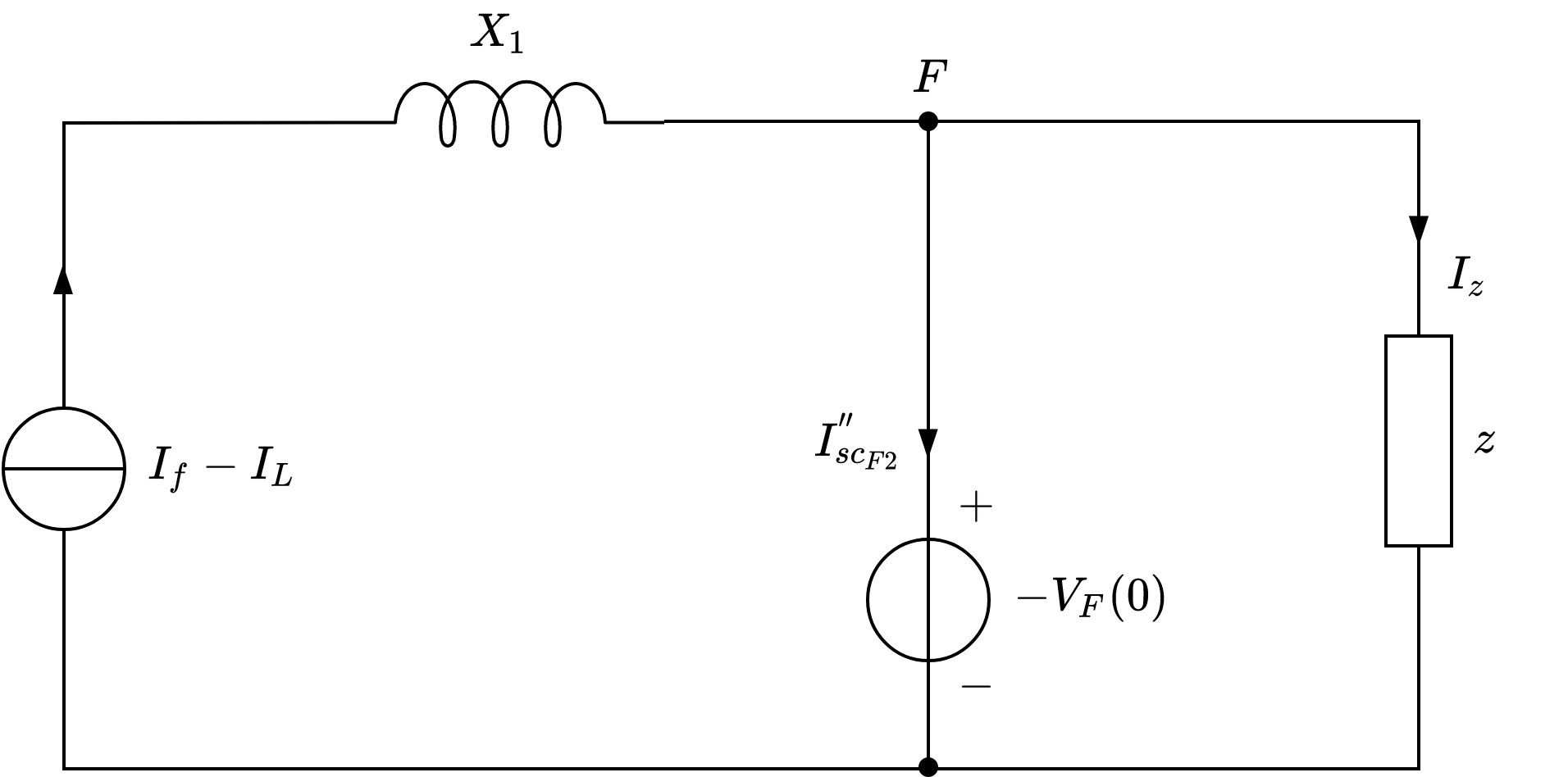}}
    \end{minipage} 
  \caption{\label{fig:VSC_SC}Equivalent circuit to calculate the SCC from IBGs: (i) post-fault condition; (ii) pre-fault condition; (iii) `pure'-fault condition.}
\end{figure}

\textcolor{black}{An analogous superposition method is adopted based on the classic SCC calculation method and the standard IEC 60909-0-2016 to quantify the SCC in a high IBG-penetrated system.} The IBGs are viewed as constant current source during a system short circuit fault as demonstrated in \cite{IEC} and \cite{NERC,7028961,ALAM2017245}. The equivalent circuit of a loaded IBG during a short circuit fault at point F is shown in Fig.~\ref{fig:VSC_SC}-(i) where $I_f$ is a pre-defined parameter representing the short circuit injection from the VSC, e.g., 1.0 p.u. and the other quantities are the same as in Fig.~\ref{fig:SG}. Applying the superposition method, the post-fault system in Fig.~\ref{fig:VSC_SC}-(i) can be viewed as the combination of pre- and `pure'- fault conditions as shown in Fig.~\ref{fig:VSC_SC}-(ii) and (iii) respectively. In the pre-fault condition, $I_L$ is the load current from the VSC and $V_F(0)$ is the bus voltage at the fault location before the fault. A voltage source with the same magnitude is applied at this point without changing the system. Different from the calculation of SCC from SGs where all the SGs in the pure-fault condition are disregarded, the inverter-based units in Fig.~\ref{fig:VSC_SC}-(iii) has to be considered as a current source with the output current being $I_f-I_L$ so that the superposition theorem holds. In order to validate the method to calculate the SCC from VSC unit, the SCC is derived from the post-fault system and the two sub-system respectively. The SCC from the IBG $c\in \mathcal{C}$ can be derived from Fig.~\ref{fig:VSC_SC}-(i) as:
\begin{equation}
\label{I_SC_VSC1}
    I_{sc_F}^{''} = I_{fc}.
\end{equation}
It can also be derived as the linear combination of $I_{sc_{F1}}^{''}$ and $I_{sc_{F2}}^{''}$, i.e.:
\begin{equation}
\label{I_SC_VSC2}
    I_{sc_F}^{''} = I_{sc_{F1}}^{''} + I_{sc_{F2}}^{''}
\end{equation}
where $ I_{sc_{F1}}^{''}$ is zero and $ I_{sc_{F2}}^{''}$ is computed by applying KCL in Fig.~\ref{fig:VSC_SC}-(ii):
\begin{equation}
\label{I_SC2_VSC}
     I_{sc_{F2}}^{''} = (I_{fc}-I_{L})-I_z.
\end{equation}
The load current in the `pure'-fault condition $I_z$ is derived from the pre-fault condition as follows:
\begin{equation}
\label{Iz}
    I_z = \frac{-V_F(0)}{z} = -I_L.
\end{equation}
Combining \eqref{I_SC_VSC2}, \eqref{I_SC2_VSC} and \eqref{Iz} gives \eqref{I_SC_VSC1}, which justifies the proposed superposition method to calculate the SCC from IBGs.

Combining the classic SCC superposition approach with the proposed model of IBG enables the SCC calculation in a general power system with both SGs and IBGs. The SCC at the fault bus $F$ can be computed through KCL in the pure-fault system where the only sources are the current at the VSC buses and the fault bus:
\begin{equation}
\label{-V_F(0)}
    -V_F(0) = \sum_{c\in \mathcal{C}} Z_{F\Phi(c)}(I_{fc}-I_{Lc})+Z_{FF}I_{sc_F}^{''}
\end{equation}
where $\Phi(c)$ maps the IBG $c\in \mathcal{C}$ to the corresponding bus index. Rearranging \eqref{-V_F(0)} yields the expression of the SCC at bus $F$ as follows:
\begin{equation}
\label{I_sc1}
    I_{sc_F}^{''} = \frac{-V_F(0)-\sum_{c\in \mathcal{C}} Z_{F\Phi(c)}(I_{fc}-I_{Lc})}{Z_{FF}}.
\end{equation}
Furthermore, the voltage $V_F(0)$ can be derived according to the pre-fault system condition:
\begin{equation}
    \label{V_F(0)}
    V_F(0) = \sum_{g\in\mathcal{G}}Z_{F\Psi(g)} I_{g}+\sum_{c\in\mathcal{C}}Z_{F\Phi(c)}I_{Lc}
\end{equation}
where $\Psi$ maps the SG $g\in\mathcal{G}$ to the index of the bus it located at. Substituting \eqref{V_F(0)} into \eqref{I_sc1} leads to:
\begin{equation}
    \label{I_sc2}
    I_{sc_F}^{''} = \frac{-\sum_{g\in\mathcal{G}}Z_{F\Psi(g)} I_{g}-\sum_{c\in \mathcal{C}} Z_{F\Phi(c)}I_{fc}}{Z_{FF}}.
\end{equation}
One can notice from \eqref{I_sc2} that pre-fault load current of the IBGs appears only in the derivation but has no impact to the final expression of the nodal SCC in the system. Based on this derivation, the SCC constraint is formulated and then linearized in the next section. \textcolor{black}{It should be noted that, although a single IBG is plotted in Fig.~\ref{fig:VSC_SC}, it can actually represent the aggregation of all the IBGs that are connected at this bus since the SCC contribution from IBGs is linear on $I_{fc}$.}

\section{SCC Constrained UC Problem Formulation} \label{sec:3}
In this section, the frequency-constrained SUC model proposed in \cite{9066910} which simultaneously optimizes the system PFR and the Synthetic Inertia (SI) from the Wind Turbines (WTs) is applied and adapted to incorporate the SCC constraint. With the linearization of the SCC constraint discussed in Section \ref{sec:3.2}, the whole problem is formulated as an MILP. \textcolor{black}{Note that only the short circuit fault at each bus in the system is considered, which is a common practice in fault analysis. Although different types of contingencies would lead to different SCC level in the system, we do not focus on the detailed contingency analysis in this paper. Instead, we take its results (minimum permissible SCC levels) as the input to the proposed model. In addition, it is possible that the SCC of some specific faults are critical given some prior knowledge about the grid, in this case, the proposed model can be extended by Norton's theorem and sequence components analysis to cope with this situation \cite{saadat2010power}.}
\subsection{Objective Function} \label{sec:3.1}
The objective of the SUC problem is to minimize the expected cost over all nodes in the given scenario tree:
\begin{equation}
    \label{eq:SUC}
    \min \sum_{n\in \mathcal{N}} \pi (n) \left( \sum_{g\in \mathcal{G}}  C_g(n) + \Delta t(n) c^s P^s(n) \right)
\end{equation}
where $\pi(n)$ is the probability of scenario $n\in \mathcal{N}$ and $C_g(n)$ is the operation cost of unit $g\in \mathcal{G}$ in scenario n including startup, no-load and marginal cost; $\Delta t(n)c^sP^s(n)$ represents the cost of the load shedding in scenario n with the three terms being the time step of scenario n, load shedding cost and shed load. 

The scenario tree is built based on user-defined quantiles of the forecasting error distribution to capture the uncertainty associated with the demand the wind generation. A rolling plan approach is implemented in the SUC model. At each time step, a 24-hour horizon SUC problem is performed with only the decisions of the current node being applied and all the future decisions discarded. At next time step, the scenario tree is updated according to the realizations of the uncertain variables and the process repeats. \textcolor{black}{Due to this rolling planning method, different realizations of nodes in the scenario tree are consistently generated at every hourly timestep according to the forecast eror quantile based probability. Therefore, the out-of-sample scheduling performances can be well captured in the simulation process rather than through Monte-Carlo simulations afterwords.}

The objective function \eqref{eq:SUC} is subjected to a number of constraints. While SCC and frequency constraints are discussed below, all other conventional UC constraints such as those of power balance, thermal units and transmission system are not listed in the paper. The readers can refer \cite{7833096} for details. 

\subsection{Formulation and Linearization of Short Circuit Current Constraint} \label{sec:3.2}
This subsection formulates the SCC constraint in the unit commitment problem. The associated quantities that influence the SCC in the system are the commitment status of the SGs and IBGs. The former can be viewed as binary decision variables i.e., $x_g,\,\forall g \in \mathcal{G}$. However, for the IBGs, their operating conditions are determined by the current available wind/solar resources rather than the system operator. Therefore, a node-dependent parameter $\alpha_c\in [0,1],\,\forall c \in \mathcal{C}$ is introduced to represent the percentage of IBGs' online capacity. The approach proposed in \cite{6672214} and \cite{7370811} is used in this paper where the online wind capacity is estimated given the current available power based on historical data.

By definition, the system admittance matrix $Y$ is formulated as follows:
\begin{equation}
    \label{Y}
    Y = Y^0 +  Y^g
\end{equation}
where $Y^0$ is the admittance matrix of the transmission lines only; $Y^g$ denotes the additional $Y$ matrix increment due to SGs' subtransient reactance. Depending on state of the SGs, the elements in $Y^g$ can be expressed as:
\begin{equation}
\label{Y2}
    Y_{ij}^g=
    \begin{cases}
    \frac{1}{X_{dg}^{''}}x_g\;\;&\mathrm{if}\,i = j \land \exists\, g\in \mathcal{G},\, \mathrm{s.t.}\,i=\Psi(g)\\
    0\;\;& \mathrm{otherwise}.
    \end{cases}
\end{equation}
The SCC constraint $I_{sc_F}^{''}\ge I_{F_{\mathrm{lim}}}^{''}$ can be derived by modifying \eqref{I_sc2}:
\begin{subequations}
\label{SCC_C}
\begin{align}
    -\sum_{g\in\mathcal{G}}Z_{F\Psi(g)} I_{g} x_g & -\sum_{c\in \mathcal{C}} Z_{F\Phi(c)}I_{fc} \alpha_c \nonumber \label{SCC_nonlinear} \\
    & \ge I_{F_{\mathrm{lim}}}^{''}Z_{FF},\,\,\forall F\in \mathcal{N}\\
    Z Y & = \mathbb{I}_N \label{ZY}
\end{align}
\end{subequations}
where $\mathbb{I}_N$ is $N$-dimensional identity matrix.  $I_g$ is computed according to \eqref{Ig} where $E_g^{''}$ is assumed to be a constant:
\begin{equation}
    E_g^{''} = \beta V_n
\end{equation}
with $V_n$ being the nominal voltage of the system and $\beta$ a parameter varying between $[0.95,1.1]$. Since the minimum initial SCC is concerned, $0.95$ is chosen in this paper.

In order to incorporate the SCC constraint into the MILP-based UC model, the nonlinearity in \eqref{SCC_C} has to be linearized. However, \eqref{ZY} is a matrix inverse constraint with binary decision variables which is difficult to deal with in an optimization problem in general. Different linearization methods are considered and compared in order to achieve a better trade-off between accuracy and computational efficiency.

\subsubsection{Analytical Method}
$N\times N$ linear equality constraints are used to enforce the relationship in \eqref{ZY}:
\begin{equation}
\label{zy}
    \sum_{k=1}^N Z_{ik}(Y_{kj}^0+Y_{kj}^g) =
    \begin{cases}
    1\;\;&\mathrm{if}\,i=j\\
    0 &\mathrm{otherwise}
    \end{cases}\;\;\;\;\;\forall i,j \in \mathcal{N}
\end{equation}
where each element in matrix $Z$ is viewed as a decision variable. Matrix $Y^g$ contains binary variables $x_g$ as in \eqref{Y2}, therefore the only nonlinearity in \eqref{zy} is the product of a continuous variable and a binary one. Take the term $\sum_k Z_{ik}Y_{k\Psi(g)}^g$ for example:
\begin{subequations}
\label{Z*x_g}
\begin{align}
    \sum_k Z_{i k}Y_{k\Psi(g)}^g & = \frac{1}{X_{dg}^{''}} \mu_g\\
    \mu_g & =Z_{i \Psi(g)}x_g. \label{Z.x}
\end{align}
\end{subequations}
The nonlinear constraint \eqref{Z.x} can be effectively linearized using McCormick Relaxation as follows:
\begin{subequations}
\label{bigM}
\begin{align}
    \mu_g & \le x_g Z_{\mathrm{max}}\\
    \mu_g & \ge -x_g Z_{\mathrm{max}}\\
    \mu_g & \le Z_{i \Psi(g)}+(1-x_g) Z_{\mathrm{max}}\\
    \mu_g & \ge Z_{i \Psi(g)}-(1-x_g) Z_{\mathrm{max}}
\end{align}
\end{subequations}
where $Z_{\mathrm{max}}$ is the upper bound of the magnitude of the elements in $Z$ matrix. All the other nonlinear terms in \eqref{SCC_nonlinear} and \eqref{zy} have the same form as \eqref{Z*x_g}. Therefore, they can be linearized using a similar approach. 

\subsubsection{Data-driven Methods}
Since \eqref{SCC_C} contains $|\mathcal{G}|$ decision variables ($x_g$) and $|\mathcal{C}|$ parameters ($\alpha_c$), the linearization of the SCC constraint can be formulated as follows:
\begin{align}
    \label{SCC_linear}
    I_{F_L}^{''} & = \sum_{g\in\mathcal{G}} k_{Fg} x_g+ \sum_{c\in\mathcal{C}} k_{Fc} \alpha_c+ \sum_{m\in\mathcal{M}} k_{Fm} \eta_m \nonumber \\
    & \ge I_{F_{\mathrm{lim}}}^{''},\,\,\forall F\in \mathcal{N}
\end{align}
where $I_{F_L}^{''}$ is the linearized SCC at Bus $F$. In order to account for the nonlinearity in \eqref{SCC_C}, the term $k_{Fm} \eta_m$ are included representing the interactions of every two SGs, which is defined as follows:
\begin{subequations}
\begin{align}
    \label{eq:x1x2}
    \eta_m  &=x_{g_1}x_{g_2},\quad \mathrm{s.t.}\{g_1,\,g_2\}=m\\
    m\in\mathcal{M} & =\{g_1,\,g_2 \mid g_1,\,g_2\in \mathcal{G}\}.
\end{align}
\end{subequations}
Note that since the 2nd order terms lead to accurate enough results as illustrated in Section \ref{sec:4.1.2}, the higher order terms are neglected in order to achieve a balance between the accuracy and computational time. The nonlinear constraint \eqref{eq:x1x2} can be then linearized using a similar way as \eqref{bigM}. The coefficients $\mathcal{K} =\{k_{Fg},\,k_{Fm},\,k_{Fc}\},\,\forall F,g,m,c$ are determined by solving the minimization problem $\forall F\in\mathcal{N}$:
\begin{subequations}
\label{DM1}
\begin{align}
    \label{min_linear_coeff}
    & \min_{\mathcal{K}} \sum_{\omega \in \Omega } \left( I_{F_L}^{''(\omega)} -I_{sc_F}^{''(\omega)} \right)^2\\
     &\left. I_{F_L}^{''(\omega)} = I_{F_L}^{''}  \right\rvert_{x_g^{(\omega)},\,\eta_m^{(\omega)},\,\alpha_c^{(\omega)}}
\end{align}
\end{subequations}
with $\omega = \{x_g^{(\omega)},\,\alpha_c^{(\omega)},\, I_{sc_F}^{''(\omega)}\}\in \Omega$ denoting the data set. It is generated by evaluating the SCC at each bus in representative system conditions. For the SGs, all the possible generator combinations are considered, whereas for the continuous parameters $\alpha_c$, there are infinite possible conditions in theory. However, according to \eqref{SCC_nonlinear}, $\alpha_c$ only influences the SCC in a linear manner as it does not appear in \eqref{ZY}. Hence, two values are sufficient to capture the linear behavior, i.e., $\alpha_c \in \{0,\,1\}$, which leads the size of $\Omega$ to $2^{|\mathcal{G}|+|\mathcal{C}|}$ in total.

Since the optimization problem \eqref{min_linear_coeff} aims to minimize the total squared error between the linearized SCC and its actual value, the conservativeness cannot be guaranteed. In order to force the relationship $I_{F_L}^{''(\omega)}\le I_{sc_F}^{''(\omega)},\,\forall \omega$, \eqref{min_linear_coeff} is modified as follows: 
\begin{subequations}
\label{DM2}
\begin{align}
    \min_{\mathcal{K}}\quad & \sum_{\omega \in\Omega} \left(I_{sc_F}^{''(\omega)} - I_{F_L}^{''(\omega)} \right)\\
    \label{coef_ctr1}
    \mathrm{s.t.}\quad & I_{F_L}^{''(\omega)}\le I_{sc_F}^{''(\omega)},\,\,\forall \omega \in \Omega.
\end{align}
\end{subequations}
With the constraint \eqref{coef_ctr1}, the linearized function is always below the real one. It is understandable that the conservativeness is achieved at a cost due to the nonlinearity. At some data points, larger errors which could have been compensated with each other in the least square optimization \eqref{min_linear_coeff} now become inevitable. As a result, there exists situation where the system actual SCC is above its permissible limit but not according to the linearized value, thus increasing the operational cost. 

Another reason for the over-conservativeness caused by \eqref{DM2} is that its objective function trys to minimize the errors of all the data points. Since the SCC limit is a hard constraint in the SUC problem, it is not essential to force the linearized SCC as close to its real value as possible everywhere. Instead, the idea of ``classification'' can be utilized to train the linear function. That is as long as the linearized SCC is on the same side of the limit as its real value, the error is of no concern. However, since the SCC is a continuous variable, which is different from the classic classification problem, some modifications need to be made in the formulation:
\begin{subequations}
\label{DM3}
\begin{align}
    \label{obj3}
    \min_{\mathcal{K}}\quad & \sum_{\omega \in \Omega_2} \left(I_{sc_F}^{''(\omega)} - I_{F_L}^{''(\omega)} \right)^2\\
    \label{coef_ctr2}
    \mathrm{s.t.}\quad & I_{F_L}^{''(\omega)}< I_{F_{\mathrm{lim}}}^{''},\,\,\forall \omega \in \Omega_1\\
    \label{coef_ctr3}
    & I_{F_L}^{''(\omega)}\ge I_{F_{\mathrm{lim}}}^{''},\,\,\forall \omega \in \Omega_3
\end{align}
\end{subequations}
where $\Omega_1 ,\, \Omega_2$ and $\Omega_3 $ are the subsets of $\Omega$, being defined as follows.
\begin{subequations}
\begin{align}
    \Omega &= \Omega_1 \bigcup\Omega_2\bigcup\Omega_3 \\
    \label{Omega1}
    \Omega_1 & = \left\{\omega\in \Omega \mid I_{sc_F}^{''(\omega)}<I_{F_{\mathrm{lim}}}^{''} \right\}\\
    \label{Omega2}
    \Omega_2 & = \left\{\omega\in \Omega \mid I_{F_{\mathrm{lim}}}^{''} \le I_{sc_F}^{''(\omega)}<I_{F_{\mathrm{lim}}}^{''} + \nu \right\}\\
    \label{Omega3}
    \Omega_3 & = \left\{\omega\in \Omega \mid I_{F_{\mathrm{lim}}}^{''} + \nu\le I_{sc_F}^{''(\omega)} \right\}.
\end{align}
\end{subequations}
Given \eqref{coef_ctr2} and\eqref{Omega1}, all the below-limit SCC can be classified correctly by the linearized function. Ideally, it is also desired to identified all the above-limit SCC. However, this may cause infeasibility due to the restricted linear structure. Therefore, a parameter $\nu\in \mathbb{R}^+$ is introduced to define $\Omega_2$ and $\Omega_3$ as in \eqref{Omega2} and \eqref{Omega3}. In this way, all the data points in $\Omega_3$ will be classified correctly and misclassification can only occur in $\Omega_2$. 

Furthermore, $\nu$ should be chosen as small as possible while ensuring the feasibility. Since it is only the data points in $\Omega_2$ whose errors are penalized in the objective function \eqref{obj3}, a good behavior could be trained, thus avoiding being over-conservative. The overall linearization performance of the three data-driven methods as well as the analytical method are compared in Section \ref{sec:4.1.2}.

\subsection{Frequency Constraints} \label{sec:3.3}
The frequency security constraints are elaborated here. A novel wind turbine SI control scheme is proposed in \cite{9066910}, which eliminates the secondary frequency dip due to WT over production and allows the WT dynamics to be analytically integrated into the system frequency dynamics. As a result, the frequency nadir constraint can be formulated as:
\begin{equation}
\label{nadir_c}
    HR\ge \frac{\Delta P_L^2T_d}{4\Delta f_\mathrm{lim}}-\frac{\Delta P_L T_d }{4} \left(D- \sum_{j\in \mathcal{F}} \gamma_j H_{s_j}^2 \right)
\end{equation}
where $H$ is the sum of conventional inertia $H_c$ and SI of all the wind farms $\sum_{j} H_{s_j}$ with $j\in \mathcal{F}$ being the set of wind farms and $R,\,T_d$ represent the system Primary Frequency Response (PFR) and its delivery time; $\Delta P,\, \Delta f_{\mathrm{lim}}$ and $D$ denote system disturbance, limit of frequency deviation and system damping. The last term can be interpreted as: SI provision from each wind farm introduces a negative damping to the system proportional to $H_{s_j}$ with the coefficient being $\gamma_j$. Furthermore, the multi-dimensional hyperboloid constraint \eqref{nadir_c} can be effectively piece-wise linearized as follows:
\begin{equation}
\label{nadir_L}
    a_i H+b_i R+\Tilde{c}_i \Tilde{H}_s+d_i \le 0, \; \forall i\in\mathcal{P},
\end{equation}
with $a_i,\,b_i$, $\Tilde{c}_i=[c_{i,1},c_{i,2},...,c_{i,|\mathcal{F}|}]$ and $d_i$ being the linearized coefficients,  $\mathcal{P}$ the set of hyperplanes and $\Tilde{H}_{s} =[H_{s_1},H_{s_2},...,H_{s_{|\mathcal{F}|}}]^\mathsf{T}$. 

Since both the frequency and SCC constraints are influenced by the SGs and IBGs, it is of interest to investigate their combined effects on system operation.

\section{Results} \label{sec:4}
Case studies on the IEEE 30- and 118-bus systems are carried out to illustrate the validity of the proposed model and the impacts of the SCC constraint on system operation as well as its interaction with system frequency constraints. \textcolor{black}{Although it may be helpful to illustrate some results based on deterministic formulation first, it turns out that the deterministic results do not present significant differences in terms of system SCC thus not being included in the paper.} The network data of the systems can be obtained in \cite{Data_system}. The MILP-base UC problem is solved by FICO Xpress on a PC with Intel(R) Core(TM) i7-7820X CPU @ 3.60GHz and RAM of 64 GB. 

\subsection{IEEE 30-Bus System} \label{sec:4.1}
\begin{figure}[!t] 
	\centering
	\vspace{-0.4cm}
	\scalebox{0.25}{\includegraphics[]{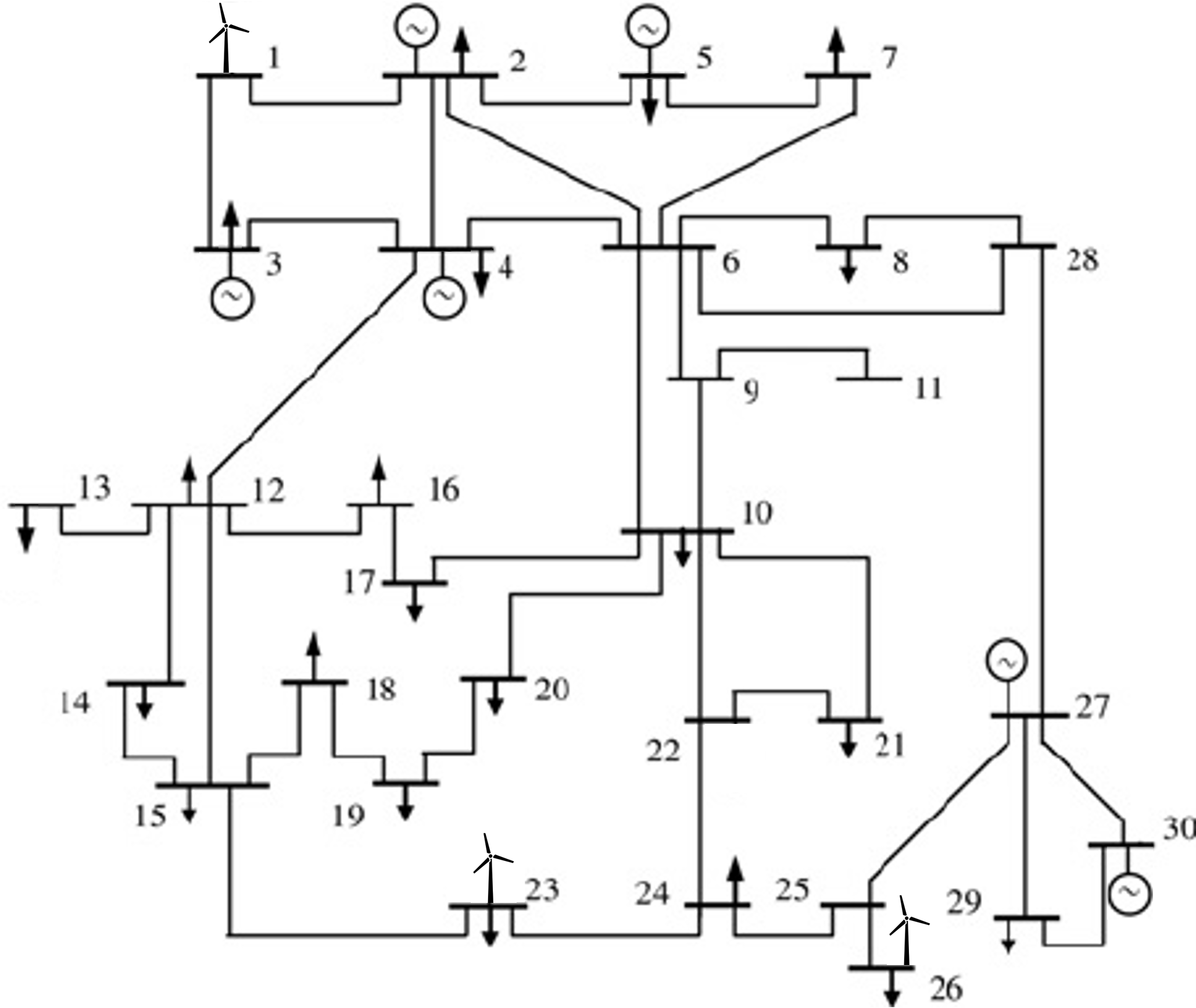}}
	\caption{Modified IEEE-30 bus system.}
	\label{fig:ieee-30}
\end{figure}

The system is modified by adding wind generation at Bus 1, 23 and 26 to model the IBG-penetrated system as shown in Fig.~\ref{fig:ieee-30}. The system parameters are set as follows: load demand $P_D \in [160,410] \,\mathrm{MW}$, damping $D = 0.5\% P_D / 1\,\mathrm{Hz}$, FR delivery time $T_d = 10\,\mathrm{s}$ and maximum power loss $\Delta P_L = 50\,\mathrm{MW}$. The frequency limits of nadir, steady state and RoCoF set by National Grid are: $\Delta f_\mathrm{lim} = 0.8\,\mathrm{Hz}$, $\Delta f_\mathrm{lim}^\mathrm{ss} = 0.5\,\mathrm{Hz}$ and $\Delta \dot f_\mathrm{lim} = 0.5\,\mathrm{Hz/s}$. The SCC from the IBGs are assumed to be $1\,\mathrm{p.u.}$ for conservativeness. 
\subsubsection{Impact of wind penetration on system SCC} \label{sec:4.1.1}
\begin{remark}
In this section, the SCC constraint is not included in the SUC model in order to study the impact of wind penetration on SCC in an unconstrained system.
\end{remark}

\begin{figure}[!t] 
	\centering
	\vspace{-0.4cm}
	\scalebox{1.2}{\includegraphics[]{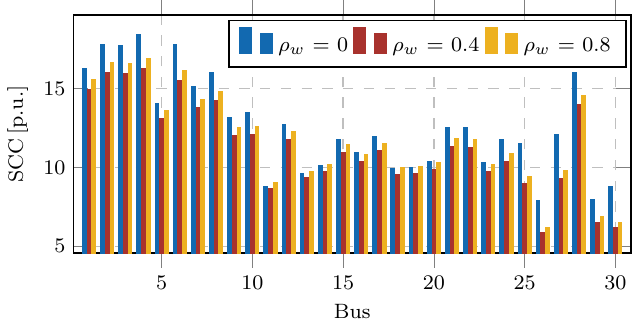}}
	\caption{SCC at different wind penetration without SI.}
	\label{fig:W_Level}
\end{figure}

\begin{figure}[!t] 
	\centering
	\vspace{-0.4cm}
	\scalebox{1.2}{\includegraphics[]{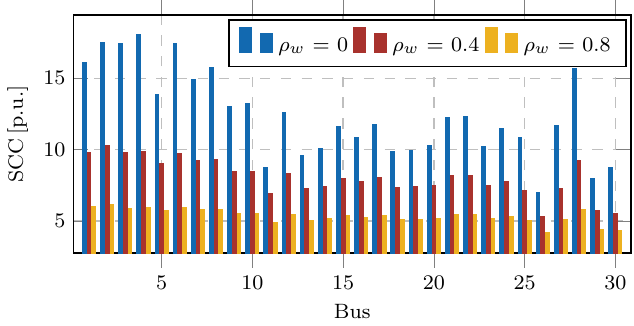}}
	\caption{SCC at different wind penetration with SI.}
	\label{fig:WL2}
\end{figure}

The SUC problem of one year is solved with total installed wind generation capacity being $800\,\mathrm{MW}$. Hours with different instantaneous wind penetration in the optimal solution are selected. The SCC at each bus is further calculated and depicted in Fig.~\ref{fig:W_Level} and Fig.~\ref{fig:WL2} where $\rho_w$ is the instantaneous wind penetration rate defined by:
\begin{equation}
    \label{eq:rho_w}
    \rho_w = \frac{P_w-P_w^{sh}}{P_D}
\end{equation}
with $P_w/P_w^{sh}$ and $P_D$ being total available/shed wind power and system demand. Note that only the cases with $\rho_w = 0,\,0.4$ and $0.8$ representing zero, medium and high wind penetration are considered for clarity. 

The situation where no SI is provided from WTs (Fig.~\ref{fig:W_Level}) is studied first. The SCC is calculated in p.u. based on $S_B =100\,\mathrm{MVA}$. For the zero wind penetration case, the SCCs in the system represented by the blue bars vary at different buses depending on their electric distance to the online conventional generators. The magnitude of the SCC in this case is the largest with an average of $12.53\,\mathrm{p.u.}$. In the medium wind penetrated system, the SCCs (red bars) at all buses decrease to some extent due to the disconnection of the conventional generators located at Bus 26 and 30. Larger reductions are therefore spotted at the nearby buses. However, as the wind penetration doubles to 80 \%, the resulted SCC even increases at each bus. This is because only one more SG at Bus 5 is disconnected whereas significant amount of SGs are forced to be online for the sole purpose of system inertia and PFR provision to maintain the frequency constraints. The resulted SCC is, in fact, dominated by the SCC boost contributed from the increased IBGs.  It is clear that in the system without SI, the reduction of SCC is limited and hence it is not necessary to include such constraints.

In the case where SI from WTs is available in the system, the SCC at each bus presents a different trend against the wind penetration. As $\rho_w$ increases to $0.4$, a remarkable SCC reduction in the range of $[21\%,\,45\%]$ is identified at all the buses. This much more significantly decreased SCC compared to that in Fig.~\ref{fig:W_Level} is due to the fact that with the SI from WTs, much fewer SGs are needed in the system to maintain the frequency constraints. For the same reason, in the high wind penetration system, the SCC at every bus further declines for another $33\%$ on average. Actually, the only synchronous generators that are online in this hour locate at Bus 2 thus resulting in the highest SCC there. The SCCs at the rest buses in the system reduce gradually as their electric distance to Bus 2 increases. Therefore, the two buses with minimum SCC are identified: Bus 26 ($4.13\,\mathrm{p.u.}$) and Bus 30 ($4.23\,\mathrm{p.u.}$). Under these conditions, it becomes critical to incorporate SCC constraints into the SUC model in order to maintain reliable system operation.
\begin{figure}[!t] 
	\centering
	\vspace{-0.4cm}
	\scalebox{1.2}{\includegraphics[]{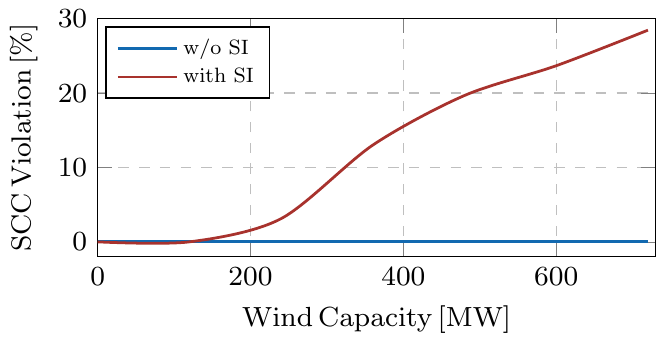}}
	\caption{SCC violation percentage at different wind penetration.}
	\label{fig:WL3}
\end{figure}

Furthermore, the SCC constraint violation in the grid during the transition towards a high wind penetration system is also investigated. The results are shown in Fig.~\ref{fig:WL3} with the minimum permissible SCC, $I_{F_{\mathrm{lim}}}^{''}$ being $5\,\mathrm{p.u.}$. Without SI from WTs, no SCC constraint violation can be found regardless of installed wind capacity. On the contrary, once the SI from WTs becomes available, an increasing proportion of operating hours will face the SCC shortage along with the higher penetration of wind generation. About $28\%$ of the total hours exhibit SCC violation at different buses at $720 \,\mathrm{MW}$ wind capacity. 

\subsubsection{Linearized SCC constraint validation} \label{sec:4.1.2}
\begin{table}[!b]
\renewcommand{\arraystretch}{1.2}
\caption{Comparison of SCC Constraint Linearization}
\label{tab:SCC_Linear}
\noindent
\centering
    \begin{minipage}{\linewidth} 
    \renewcommand\footnoterule{\vspace*{-5pt}} 
    \begin{center}
        \begin{tabular}{ c | c | c | c | c | c | c }
            \toprule
             \multirow{2}{2.8em}{\textbf{Method}} & \multicolumn{2}{c|}{\textbf{Type I Error}} &\multicolumn{2}{c|}{\textbf{Type II Error}}  & \multirow{2}{2.9em}{\textbf{ Cost}\\$[\mathrm{k\pounds}/h]$} & \multirow{2}{2.6em}{\textbf{\,\,Time}\\$[\mathrm{s/step}]$}\\ 
            \cline{2-5}
            &$N_e$ & $err$ & $N_e$ & $err$ & \\ 
            \cline{1-7} 
            AM & $0$ & $0$ & $0$ & $0$ & $28.60$& $>1e3$ \\
            \cline{1-7} 
            DM-1 & $18$ & $1.18\%$ & $213$ & $-2.60\%$ & $28.92$& $1.51$  \\
            \cline{1-7} 
            DM-2 & $0$ & $0$ & $7379$ & $-16.27\%$ & $32.25$ & $3.41$  \\
            \cline{1-7} 
            DM-3 & $0$ & $0$ & $13$ & $-0.45\%$ & $28.60$ & $3.78$ \\
           \bottomrule
        \end{tabular}
        \end{center}
    \end{minipage}
\end{table} 
Before incorporating the SCC constraint into the SUC problem, the performance of the linearized SCC constraints proposed in Section \ref{sec:3.2} is examined first. The results are shown in Table~ \ref{tab:SCC_Linear} where AM and DM-1/2/3 denote the analytical method based on matrix inverse \eqref{zy} and the three data-driven methods according to \eqref{DM1} - \eqref{DM3} respectively. Two types of error are analyzed with Type I being the SCC which satisfies the constraint according to the linearized value but actually violates and Type II being the opposite. $N_e$ and $err$ are the total number and the averaged value of the errors defined as follows:
\begin{subequations}
\begin{align}
    N_e &  = |\mathcal{E}|\\
    err &  = \frac{1}{N_e}\sum_\mathcal{E} \frac{ I_{F_L}^{''(\mathcal{E})}-I_{sc_F}^{''(\mathcal{E})}}{I_{sc_F}^{''(\mathcal{E})}}
\end{align}    
\end{subequations}
with $\mathcal{E}$ being the set of errors. For the analytical method, no error is observed since there is no approximation in the formulation. However, the computational time is extremely long due to the presence of the matrix inverse with decision variables \eqref{zy} - \eqref{bigM}, which limits its practical application in large scale systems. This challenge is resolved by the data-driven methods with much shorter computation time. The least square minimization is used in DM-1. It can be seen that both types of the error occur with relatively small $err$. As Type-I error leads to insecure operation point, \eqref{coef_ctr1} is added in DM-2 to eliminate such error. The effectiveness is manifested by zero Type-I error in the table. However, an averaged Type-II error of $-16.27\%$ and an operational cost increase of $13\%$ demonstrate its over-conservativeness in some operating conditions. DM-3 adapts the idea of classification in the optimization, which only minimizes the error around the limit. The results are promising with few Type II errors and negligible $err$. These errors almost have no impact on the system operation cost compared with the analytical method. Although the time is slightly longer than DM-2, it is still acceptable and represents significant reduction over AM. As a result, DM-3 is used for the rest of the study.

\subsubsection{SCC constrained UC problem} \label{sec:4.1.3}
\begin{figure}[!t] 
	\centering
	\vspace{-0.4cm}
	\scalebox{1.2}{\includegraphics[]{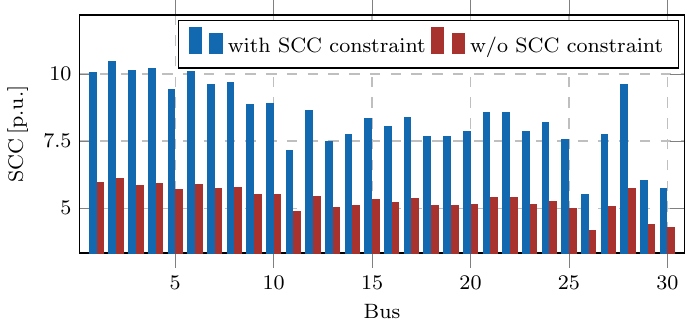}}
	\caption{(Un)constrained SCC in the system.}
	\label{fig:SCC_C}
\end{figure}
\begin{table}[!b]
\renewcommand{\arraystretch}{1.4}
\caption{Assessment of SCC Constraint Level}
\label{tab:SCC_C}
\noindent
\centering
    \begin{minipage}{\linewidth} 
    \renewcommand\footnoterule{\vspace*{-5pt}} 
    \begin{center}
        \begin{tabular}{ c | c | c | c | c | c | c | c | c  }
            \toprule
             \multirow{2}{2.2em}{$I_{F_\mathrm{lim}}^{''}$\\ $[\mathrm{p.u.}]$} & \multicolumn{2}{c|}{\textbf{min SCC}} &\multicolumn{5}{c|}{\textbf{Online SG Number}}  & \multirow{2}{2.5em}{\textbf{   Cost}\\$[\mathrm{k\pounds}/h]$} \\ 
            \cline{2-8}
            & p.u. & Bus & $N_2$  & $N_{3}$  & $N_{4,5}$  & $N_{26}$ & $N_{30}$ \\ 
            \cline{1-9} 
            $0.00$ & $4.13$ & 26 & $2$ & $0$ & $0$ & $0$ & $0$ &$10.60$\\
            \cline{1-9} 
            $5.00$ & $5.49$ & 26 & $3$ & $1$  & $0$ & $0$ & $0$ &$21.02$\\
            \cline{1-9} 
            $6.00$ & $6.19$ & 26 & $3$ & $0$ & $0$ & $1$ & $1$ &$26.41$\\
            \cline{1-9} 
            $7.00$ & $7.14$ & 29 & $3$ & $1$ & $0$ & $2$ & $1$ & $37.74$\\
           \bottomrule
        \end{tabular}
        \end{center}
    \end{minipage}
\end{table} 
In this section, the SCC constraint is incorporated into the SUC model to evaluate its effectiveness as well as the influence on system operation. The SCC at each bus in the system with and without the SCC constraint is shown in Fig.~\ref{fig:SCC_C}. The instantaneous wind penetration rate in the system is $0.8$ and the minimum permissible SCC in this case is set to be 5 $\mathrm{p.u.}$, i.e., $I_{F_{\mathrm{lim}}}^{''} = 5,\,\,\forall F\in \mathcal{N}$ such that a short circuit ratio of 3 (strong grid) can be maintained. \textcolor{black}{It should be noted that depending on the type of bus that is of concern, different strategies shall be considered to determine this minimum permissible SCC level. Specifically, for the point of common coupling of IBGs, a stiff voltage signal is desired to ensure the synchronization of PLLs and the stability of current grid-following converters. This is often indicated by a system strength (short circuit ratio) of 3 or more \cite{Jia18}, which would lead to an appropriate SCC constraint level at this point given the knowledge of IBGs' capacity. For other buses, the SCC constraint level may be decided by the settings of critical protection devices, the requirement of voltage stability margin. In general, the determination of the minimum SCC level can be very complex \cite{AEMO} and should be conducted before the system scheduling process thus being out of the scope of the problem at hand.}

If not being constrained, the SCC at some buses in the system is below 5 p.u. as illustrated by the red bars in Fig.~\ref{fig:SCC_C}. Once the SCC constraint is included, the SCC at all buses is above the limit, which demonstrates the effectiveness of the proposed method. This is achieved by increasing the online SGs to provide the SCC during faults.

Different levels of the SCC limit are also considered to better review their impacts on generator dispatch and system operation cost. The results are shown in Table~\ref{tab:SCC_C} where the base case ($I_{F_\mathrm{lim}}^{''}=0$) corresponds to the red bars in Fig.~\ref{fig:SCC_C} and $N_i$ represents the online SG number at bus $i$. As the level of the SCC limit increases from $5\,\mathrm{p.u.}$ (Case A) to $7\,\mathrm{p.u.}$ (Case C), the actual minimum SCC in the system is always higher than the limit and the bus at which it locates also changes. It is understandable that higher SCC in the system would lead to larger system operation cost. For example, the cost in Case A is about two times that of the base case since the number of online SGs doubles. It is also worth noting that in Base B, the SGs at Bus 26 and 30 are connected to the grid to replace the SG at Bus 3. Although the SG at Bus 3 has lower operating cost, the SGs at Bus 26 and 30 are more effective in supplying SCC for the limiting bus, demonstrating the spacial phenomena of the SCC constraints. Moreover, the minimum SCC in the system is changed from Bus 26 to 19 in Case C because of the specific SG combination, which illustrates that only confining certain bus with minimum SCC may not be sufficient during some operating conditions.

\subsubsection{Assessment of IBGs' PE overcapacity}
\begin{figure}[!t] 
	\centering
	\vspace{-0.4cm}
	\scalebox{1.2}{\includegraphics[]{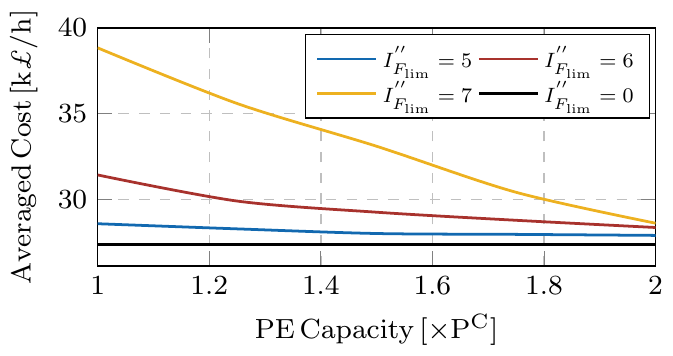}}
	\caption{System operation cost against PE capacity.}
	\label{fig:OverC}
\end{figure}

Instead of dispatching more SGs, the SCC in the system can be improved by increasing the PE capacity of the IBGs as well, through either increasing the capacity of the semiconductor devices or utilizing advanced technologies \cite{9025185}. In this section, the value of IBGs' short-term PE overcapacity is investigated. The results are depicted in Fig.~\ref{fig:OverC} where the horizontal axis represents the ratio to the base PE capacity $P^C$. The averaged system operation cost is reduced by increasing the PE capacity due to the additional SCC capability from IGBs. Moreover, as the SCC limit becomes higher, the IBGs' overcapacity is more valuable. However, the marginal value of overcapacity quickly reduces as it approaches to the unconstrained case. This results and further studies can be used to inform the optimal sizing of IBGs' PE capacity.

\subsection{IEEE 118-bus system}
The scalability of the proposed method is demonstrated through IEEE 118-bus system with the demand $P_D\in[2000,5600]\,\mathrm{MW}$, largest loss of generation $\Delta P_L = 500\,\mathrm{MW}$ and $S_B = 1000\,\mathrm{MVA}$. Wind generation is added at Bus 3, 41, 72 and 87 and the parameters for the frequency regulation remains unchanged. 

\begin{figure}[!t] 
	\centering
	\vspace{-0.4cm}
	\scalebox{1.2}{\includegraphics[]{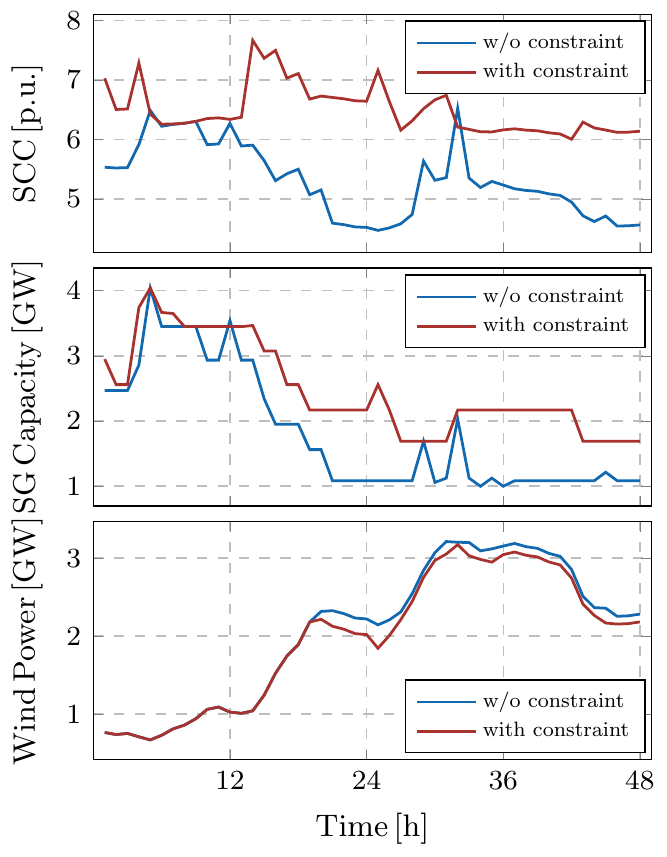}}
	\caption{SUC results within two days: (i) SCC; (ii) online SG capacity; (iii) Wind generation.}
	\label{fig:SCC_Time}
\end{figure}

The SUC problem is performed in different conditions to compare the effects of the SCC constraint and the IBGs' overcapacity. The results are listed in Table.~\ref{tab:118} with the base size of PE capacity $P^C = 8000\,\mathrm{MW}$. Nearly two thirds of the total hours present shortage of SCC without SCC constraint. This number reduces significantly to $18.97\%$ if the total PE capacity of IBGs is increased by $50\%$. Once the SCC is constrained in the SUC model, the violation disappears with an increased averaged cost of $11.6\%$. The overcapacity in this case decreases the system operation cost since more SCC is provided by the IBGs thus reducing the committed SGs during high wind penetration period. In addition, there is a clear increase in the computational time when SCC constraints are added but it is still in a manageable range. In fact, it can be also noted when overloading capability of IBGs increases, the computational time decreases as fewer SCC constraints are active. Moreover, an example of SCC variation during 48 hours at Bus 87 is plotted in Fig.~\ref{fig:SCC_Time}-(i) with the PE capacity being $P^C$ and $I_{F_{\mathrm{lim}}}^{''}=6\,\mathrm{p.u.}$. Without the constraint, SCC violation is observed most of the time (blue curve), which is successfully eliminated by the SCC constraint (red curve). This is achieved by increasing the online SGs as in  Fig.~\ref{fig:SCC_Time}-(ii) where more SGs are dispatched during the hours with SCC constraint violation. \textcolor{black}{In details, the SGs that will be switched on depend on the combination effects of the SG capacity, marginal cost and its electric distance to the critical bus. If a large SCC shortage occurs, e.g., Hour 24, the closest SG at Bus 87 is dispatched regardless of its higher marginal cost; On the contrary, if the SCC shortage is small (Hour 36), farther SGs located at Bus 69 may be switched on to reduce the overall operation cost}. Furthermore, \textcolor{black}{because of these additionally dispatched SGs, less wind power is utilized during the time $\mathrm{[20,48]\,h}$ when there is more than enough wind resource in the system as illustrated in Fig.~\ref{fig:SCC_Time}-(iii). Unsurprisingly, larger SCC shortage is likely to lead to more SGs being online thus more wind power being curtailed (red curve), which also justifies the cost increase in Table III.}

\begin{table}[!t]
\renewcommand{\arraystretch}{1.2}
\caption{SCC comparison in IEEE 118-bus system}
\label{tab:118}
\noindent
\centering
    \begin{minipage}{\linewidth} 
    \renewcommand\footnoterule{\vspace*{-5pt}} 
    \begin{center}
        \begin{tabular}{c | c || c | c | c }
            \toprule
            \multirow{2}{4.2em}{\textbf{\,\,\,\; SCC\\Constraint} } & \multirow{2}{4.2em}{\textbf{IBGs' PE\\Capacity} }& \multirow{2}{4em}{\textbf{ \, SCC\\Violation} } & \multirow{2}{2.8em}{\textbf{ Cost}\\$[\mathrm{k\pounds/h}]$} & \multirow{2}{3em}{\textbf{\,\,\,Time}\\ $[\mathrm{s/step}]$} \\ 
            & & & &\\
            \cline{1-5}
            \multirow{2}{2.8em}{w.o.} & $P^C$ & $67.93$\% & $56.71$ & $19.55$  \\ 
            \cline{2-5} 
            & $1.5 P^C$ & $18.97\%$ & $-$ & $-$\\
            \cline{1-5} 
            \multirow{2}{2.8em}{with} & $P^C$ & $0$ & $63.31$ & $32.32$  \\
            \cline{2-5} 
            & $1.5P^C$ & $0$ & $58.78$ & $23.04$ \\
           \bottomrule
        \end{tabular}
    \end{center}
    \end{minipage}
\end{table} 

\section{Conclusion}    \label{sec:5}
A short circuit current constrained SUC model is proposed in this paper in order to deal with the SCC shortfalls due to the increasing penetration of IBGs. The SCC from SGs as well as IBGs is modeled in the formulation leading to a highly nonlinear constraint, which is effectively linearized based on the data obtained offline. It is validated that the linearized constraint provides almost the same accuracy as the nonlinear version but consumes much less computational time. 

The influence of the SCC constraint on the system operation and its interaction with the frequency constraints are thoroughly investigated via IEEE 30-bus system. The results demonstrate that the SCC in the system declines dramatically as the wind penetration increases if SI is available from WTs. Overcapacity of the IBGs helps to improve system SCC but needs more investment cost. Employed with the SCC constraint, the system SCC limit can be maintained with higher operational cost depending on the wind penetration. The scalability of the proposed model is assessed through IEEE 118-bus system.
\vspace{-0.25cm}
\bibliographystyle{IEEEtran}
\bibliography{bibliography}




\vfill

\enlargethispage{-50mm}

\end{document}


\begin{varwidth}{\linewidth}

\begin{tikzpicture}
\begin{axis}[
    ybar=0.8pt,
    bar width=.02cm,
    width=7.25cm,
    height=4cm,
    symbolic x coords={1,2,3,4,5,6,7,8,9,10,11,12,13,14,15,16,17,18,19,20,21,22,23,24,25,26,27,28,29,30},
    enlarge x limits=0.03,
    xlabel={$\mathrm{Bus}$},
    ylabel={$\mathrm{SCC}\,\mathrm{[p.u.]}$},
    xtick={5,10,15,20,25,30},
    xmajorgrids=true,
    ymajorgrids=true,
    legend style={at={(0.635,0.982)}, anchor=north,legend columns=-1}, legend cell align={left},
    grid style=dashed,nodes={scale=0.75, transform shape}]
\footnotesize
\addplot[
    thick,
    color=pBlue,
    fill=pBlue, 
    ]
    table {data/WL_SI/data1.txt};    
    \addlegendentry{\footnotesize $\rho_w$ = 0}        
\addplot[
    thick,
    color=pRed,
    fill=pRed, 
    ]
    table {data/WL_SI/data2.txt};        
    \addlegendentry{\footnotesize $\rho_w$ = 0.4}   
    
\addplot[
    thick,
    color=pYellow,
    fill=pYellow, 
    ]
    table {data/WL_SI/data3.txt};        
    \addlegendentry{\footnotesize $\rho_w$ = 0.8}        

\end{axis}

\end{tikzpicture} 

\end{varwidth}